\documentclass[prl,twocolumn,aps,showpacs,preprintnumbers]{revtex4}
\usepackage{epsfig,epsf,psfrag}
\usepackage{graphicx}
\usepackage{epic,eepic}
\usepackage{color,pstcol}
\begin{document}
\title{Incoherence induced sign change in noise cross-correlations: A case study in the full counting statistics of a pure spin pump}
\date{\today}
\author{Colin Benjamin}
\affiliation{Quantum Information Group, School of Physics and
Astronomy, University of Leeds, Woodhouse Lane, Leeds LS2 9JT,
UK.}
\begin{abstract}
The full counting statistics of a non adiabatic pure spin pump are calculated with particular emphasis on the second and third moments. We show that incoherence can change the sign of spin shot noise cross-correlations from negative to positive, implying entanglement for spin-singlet electronic sources, a truly counterintuitive result. The third moment on the other hand is shown to be much more resilient to incoherence.
\end{abstract}
\maketitle
{\em Introduction}: Charge or spin transport is a statistical process involving
electrons carrying definite amounts of spin or charge, since
charge or spin current fluctuates in time. Therefore, in addition
to knowing the mean charge or spin current passing through a normal conductor one needs to know the noise as well as the other transport moments in
order to fully characterize charge or spin electron motion. To do this one takes recourse to the full counting statistics(FCS), which gives us the
complete knowledge about all the moments of the distribution of
the number of transferred charges or spins. The full counting statistics
for a non-adiabatic pure spin pump is analyzed in both the completely coherent and incoherent regimes.

Shot noise cross-correlations, the second moment, in solid state devices have been studied for a long time. Some of these studies include normal metal-superconductor hybrid structures\cite{martin}, coulomb blockaded quantum dots\cite{cottet}, exploiting the Rashba scattering\cite{egues}, etc. However, an experimental demonstration has thus far been lacking. This is mainly due to the difficulty in controlling environmental effects like incoherence. It begs the question how to deal with incoherence and reduce it. In this work a novel scheme is proposed in which the incoherence present in such systems can be used as a resource. We particularly concentrate on the electronic spin. The reason for dwelling on the spin instead of charge is because there have been many works on the charge counting statistics however works on the full counting statistics for spin are less visible. However, they have been attempted in different context to that which is the topic of this rapid communication. For example, in [\onlinecite{nazarov}], the FCS of spin currents was first attempted, the FCS of spin transfer through ultrasmall quantum dots in context of Kondo effect was attempted in [\onlinecite{komnik}] while in [\onlinecite{urban}] a study of FCS in interacting quantum dots attached to ferromagnetic leads revealed super-poissonian transport. Many works revolve around the spin shot-noise cross-correlations. Among the notable works on spin shot-noise cross-correlations mention may be made of: spin current shot noise of (i) a single quantum dot coupled to an optical microcavity and a quantized cavity field\cite{djuric}, (ii) a realistic superconductor-quantum dot entangler\cite{sauret}, and (iii) a spin transistor\cite{he}. Positive spin shot noise cross-correlations for spin-singlet electronic sources could be a signature of entanglement\cite{been} too. In this letter the properties of the third moment are also calculated. The reason for looking at the third moment is because the third moment is predicted to be much more resilient to incoherence\cite{jc-been}. In our work we prove this statement by an exact analytical calculation.

In this letter we find that in the coherent transport regime the current and spin shot noise cross-correlations are similar to that in Ref.\cite{bingdong}. The effect of incoherence on odd moments is negligible. The current
and the third moment do not change much with incoherence. In contrast the second moment, i.e.,  spin shot noise becomes completely positive. An extremely counter-intuitive result. For the third moment spin auto or cross-correlations do not change much from the coherent and incoherent transport regimes. This shows the resilience of the third moment to incoherence. The main body of this letter starts with an explanation of the model. The coherent density matrix equation is then analyzed separate from the incoherent density matrix equation to bring out the differences. Lastly we bring out a perspective on future endeavors.

\begin{figure}[h]
\centerline{
\includegraphics[width=7cm,height=3cm]{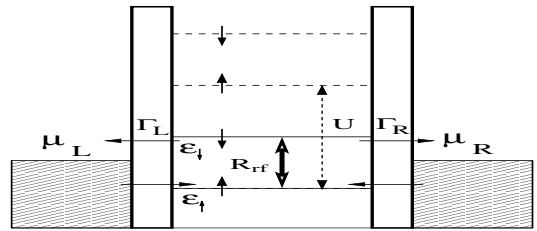}}
\caption{The model system.}
\end{figure}
{\em Model}: The model of Ref.\cite{bingdong} is the
starting point. It is depicted in Fig.1. The single electron
levels in the dot are split by an external magnetic field  $B$.
$\epsilon_{\uparrow}-\epsilon_{\downarrow}=g_{z}\mu_{B}B=\Delta$(Zeeman
energy), where $g_z$ is effective electron g-factor in z-direction
and $\mu_B$ is Bohr magneton. No bias voltage is applied across
the leads. An additional oscillating magnetic field
$B_{rf}(t)=(B_{rf}\cos(\omega t),B_{rf}\sin(\omega t))$ applied
perpendicularly to constant field B with frequency $\omega$ nearly
equal to $\Delta$ can pump the electron to higher level where its
spin is flipped, then the spin down electron can tunnel out of the
leads. Coulomb interaction in the quantum dot is considered to be strong
enough to prohibit double occupation. No extra electrons can enter
the quantum dot before the spin-down electron exits. The
Hamiltonian of ESR induced spin battery under consideration is written as:
\begin{eqnarray}
\label{eq_H}
H&=&\sum_{\eta,k,\sigma}\epsilon_{\eta,k,\sigma}c_{\eta,k,\sigma}^{\dagger}c_{\eta,k,\sigma}+
\sum_{\sigma}\epsilon_{\sigma}c_{d\sigma}^{\dagger}c_{d\sigma}+
Un_{d\uparrow}n_{d\downarrow}\nonumber\\
&+&\sum_{\eta,k,\sigma}(V_{\eta}c_{\eta,k,\sigma}^{\dagger}c_{d\sigma}+h.c.)+H_{rf}(t)
\end{eqnarray}
In the above equation,
$c_{\eta,k,\sigma}^{\dagger}(c_{\eta,k,\sigma})$ and
$c_{d\sigma}^{\dagger}(c_{d\sigma})$ are the creation and
annihilation operators for electrons with momentum $k$, spin
$\sigma$ and energy $\epsilon_{\eta,k,\sigma}$ in lead
$\eta(=L,R)$ and for spin $\sigma$ electron on the quantum dot.
The third term describes coulomb interaction among electrons on
the quantum dot. The fourth term describes tunnel coupling between
quantum dot and reservoirs. $H_{rf}(t)$ describes the coupling
between the spin states due to the rotating field $B_{rf}(t)$ and
can be written in rotating wave approximation as:
\begin{equation}
\label{eq_Href}
H_{rf}(t)=R_{rf}(c_{d\uparrow}^{\dagger}c_{d\downarrow}e^{i \omega
t }+ c_{d\downarrow}^{\dagger}c_{d\uparrow}e^{-i \omega t })
\end{equation}
with, ESR rabi frequency $R_{rf}=g_{\perp}\mu_{B}B_{rf}/2$, with
$g$-factor $g_{\perp}$ and amplitude of rf field $B_{rf}$.

The quantum rate equations for the density matrix can be easily
derived as in Ref.\cite{bingdong}. $\rho_{00}$ and $\rho_{\sigma
\sigma}$ describe occupation probability in QD being respectively
unoccupied and spin-$\sigma$ states and off-diagonal term
$\rho_{\uparrow \downarrow (\downarrow \uparrow)}$ denotes
coherent superposition of two coupled spin states in quantum dot.
The doubly occupied is prohibited due to infinite coulomb
interaction $U\rightarrow \infty$. To derive the density matrix,
we proceed as follows. The time dependence can be removed from
Eqs. [\ref{eq_H}-\ref{eq_Href}], by using the following unitary
transformation\cite{zhang}:
\begin{equation}
\label{eq_U}
U=e^{\frac{-i\omega
t}{2}[\sum_{k,\eta}(c_{\eta,k,\downarrow}^{\dagger}c_{\eta,k,\downarrow}^{}-c_{\eta,k,\uparrow}^{\dagger}c_{\eta,k,\uparrow}^{})+(c_{d,\downarrow}^{\dagger}c_{d,\downarrow}^{}-c_{d,\uparrow}^{\dagger}c_{d,\uparrow}^{})]}
\end{equation}
The Hamiltonian is then redefined in the rotating reference form as follows:
\begin{eqnarray}
\label{H_rot} H_{RF}&=&U^{-1}HU+i\frac{dU^{-1}}{dt}U\nonumber\\
&=&\sum_{\eta,k,\sigma}\bar\epsilon_{\eta,k,\sigma}c_{\eta,k,\sigma}^{\dagger}c_{\eta,k,\sigma}+
\sum_{\sigma}\bar\epsilon_{\sigma}c_{d\sigma}^{\dagger}c_{d\sigma}+Un_{d\uparrow}n_{d\downarrow}\nonumber\\
&+&\sum_{\eta,k,\sigma}(V_{\eta}c_{\eta,k,\sigma}^{\dagger}c_{d\sigma}+h.c.)+R_{rf}(c_{d\uparrow}^{\dagger}c_{d\downarrow}+c_{d\downarrow}^{\dagger}c_{d\uparrow})\nonumber\\
\end{eqnarray}
In the above equation,
$\bar\epsilon_{\uparrow}=\epsilon_{D}-\frac{\Delta}{2}+\frac{w}{2}$,
and $\bar\epsilon_{\downarrow}=\epsilon_{D}+\frac{\Delta}{2}-\frac{w}{2}$,
while $\bar\epsilon_{\eta k \uparrow}=\epsilon_{\eta k
}+\frac{w}{2}$ and $\bar\epsilon_{\eta k \downarrow}=\epsilon_{\eta k \downarrow}-\frac{w}{2}$.

 To get the density matrix from the above Hamiltonian, the
following procedure is used. An electron operator affecting only the electron on the dot can be written in terms of $|p><p|, p=0, \uparrow,
\downarrow$. Writing, for the annihilation operator of the
dot $c_{d\sigma}=|0><\sigma|$, and for the creation operator for
the dot $c_{d\sigma}^{\dagger}=|\sigma><0|$, the
Hamiltonian is rewritten in terms of the three states: $|0>, |\uparrow>,
|\downarrow>$, corresponding to empty state, a single electron
with spin-up and single electron with spin-down. The doubly
occupied state in the dot is prohibited by the fact that U is
taken to be extremely large. Thus the Hamiltonian reduces to:
\begin{eqnarray}
H&=&\sum_{\eta,k,\sigma}\bar\epsilon_{\eta,k,\sigma}c_{\eta,k,\sigma}^{\dagger}c_{\eta,k,\sigma}+
\sum_{\sigma}\bar\epsilon_{\sigma}|\sigma><\sigma|+Un_{d\uparrow}n_{d\downarrow}\nonumber\\
&+&\sum_{\eta,k,\sigma}(V_{\eta}c_{\eta,k,\sigma}^{\dagger}|0><\sigma|+h.c.)+R_{rf}(|\uparrow><\downarrow|\nonumber\\
&+&|\downarrow><\uparrow|)
\end{eqnarray}
The elements of the density matrix $\rho_{mn}$ in dot spin basis
are expectation values of operators $|n><m|$, with
$n,m=0,\uparrow,\downarrow$, so we can write-$\rho_{00}=<|0><0|>,
\rho_{\sigma\sigma}=<|\sigma><\sigma|>,
\rho_{\sigma\bar\sigma}=<|\bar\sigma><\sigma|> $.
The time evolution of the density matrix elements can be expressed
in terms of expectation values for new operators\cite{thesis_pedersen}. For instance,
\begin{eqnarray}
\label{eq_rho00} i\dot\rho_{00}&=&i \frac{\partial}{\partial
t}<|0><0|>=<[|0><0|,H]>\nonumber\\
\dot\rho_{00}&=&i[H|0><0|-|0><0|H], \nonumber\\
 &=& i[V_{\eta}^{*}|\sigma><0|c_{\eta k \sigma}-V_{\eta}|0><\sigma|c_{\eta k
 \sigma}^{\dagger}],\nonumber\\
&=&[V_{\eta}^{*}G_{\eta k \sigma}^{<}(t,t)-V_{\eta}G_{0 \sigma,
\eta k \sigma }^{<}(t,t)]
\end{eqnarray}
The approximated current Green's functions are (using Ref.\cite{thesis_pedersen}) as a guide we have:
\begin{eqnarray}
\label{eq_Gapprx} \small G_{0\sigma,\eta k
\sigma'}^{<}(t,t')&=&\int dt_{1}[G_{0\sigma
\sigma'}^{R}(t,t_{1})V_{\eta k \sigma'}^{*}g_{\eta k
\sigma'}^{<}(t_{1},t')\nonumber\\
&+&G_{0\sigma \sigma'}^{<}(t,t_{1})V_{\eta k
\sigma'}^{*}g_{\eta k \sigma'}^{A}(t_{1},t')],\nonumber\\
G_{\eta k \sigma',0 \sigma}^{<}(t,t')&=&\int dt_{1}[g_{\eta k
\sigma'}^{R}(t,t_{1})V_{\eta k \sigma'}G_{0 \sigma'
\sigma}^{<}(t_{1},t')\nonumber\\
 &+&g_{\eta k
\sigma'}^{<}(t_{1},t')V_{\eta k \sigma'}G_{0 \sigma'
\sigma}^{A}(t_{1},t')]
\end{eqnarray}
The $G_{0\sigma\sigma'}$'s are the green functions for the dot,
while $g_{\eta k \sigma}$ is the Green's function for the
$\eta$-lead in absence of tunnelling.

From the convolution theorem for Fourier transforms,
\begin{equation}
\int dt_{1}A(t-t_{1})B(t_{1}-t)=\int du A(u)B(-u)=\int
\frac{dw}{2\pi}A(w)B(w).
\end{equation}
Inserting the approximated current Green's functions from
Eqs.\ref{eq_Gapprx} into Eq.\ref{eq_rho00} and Fourier
transforming one gets:
\begin{eqnarray}
\dot \rho_{00}&=&|V_{\eta}|^{2}[G_{0\sigma\sigma}^{<}(w)(g_{\eta k
\sigma}^{R}(w)-g_{\eta k \sigma}^{A}(w))\nonumber\\
 &+&g_{\eta k
\sigma}^{<}(w)(G_{0 \sigma \sigma}^{A}(w)-G_{0 \sigma
\sigma}^{R}(w))]
\end{eqnarray}
The general property for Green's functions
$G^{>}-G^{<}\equiv G^{R}-G^{A}$, is then used-
\begin{eqnarray}
\dot \rho_{00}&=&|V_{\eta}|^{2}[G_{0\sigma\sigma}^{<}(w)(g_{\eta k
\sigma}^{>}(w)-g_{\eta k \sigma}^{<}(w))\nonumber\\ &+&g_{\eta k
\sigma}^{<}(w)(G_{0 \sigma \sigma}^{<}(w)-G_{0 \sigma
\sigma}^{>}(w))]
\end{eqnarray}
The lesser Green's function then becomes-
\begin{eqnarray}
g_{\eta k \sigma}^{<}(t)&\equiv& <c_{\eta k
\sigma}^{\dagger}c_{\eta k \sigma }(t)>=i e^{-i \epsilon_{\eta k
\sigma}t}<c_{\eta k
\sigma}^{\dagger}c_{\eta k \sigma}>\nonumber\\
&=&i e^{-i \epsilon_{\eta k \sigma}t} f_{\eta}(\epsilon_{\eta k
\sigma}),
\end{eqnarray}
where, $f(\epsilon)$ is the Fermi function. Performing a fourier
transformation yields
\begin{eqnarray}
g_{\eta k \sigma}^{<}(w)&=&2\pi i f_{\eta}(\epsilon_{\eta k
\sigma})\delta(w-\epsilon_{\eta k \sigma}),\mbox{ and similarly}\nonumber\\
 g_{\eta k \sigma}^{>}(w)&=&-2\pi i
[1-f_{\eta}(\epsilon_{\eta k \sigma})]\delta(w-\epsilon_{\eta k
\sigma})
\end{eqnarray}
Substituting the above expressions in Eqs.\ref{eq_rho00}, and
using the coupling parameter $\Gamma_{\sigma}^{\eta}(\epsilon)=2\pi\sum_{k}|V_{\eta}|^{2}\delta(\epsilon-\epsilon_{\eta
k \sigma })$ gives-
\begin{equation}
\label{eq_rho_oo_new}
 \dot \rho_{00}=\frac{-i}{2\pi}\int dw\sum_{\eta
\sigma}\{\Gamma_{\sigma}^{\eta}(1-f_{\eta}(w))G_{0\sigma\sigma}^{<}(w)+\Gamma_{\sigma}^{\eta}f_{\eta}(w)G_{0\sigma\sigma}^{>}(w)
\}
\end{equation}
The lesser and greater Greens functions for the dot can be derived
using the same formalism as in Ref.\cite{thesis_pedersen}. Thus,
$G_{0\sigma\sigma}^{<}(w)=2\pi i
\rho_{\sigma\sigma}\delta(w-\epsilon_{\sigma})$, and
$G_{0\sigma\sigma}^{>}(w)=-2\pi i
\rho_{00}\delta(w-\epsilon_{\sigma})$. After substituting these
expressions in Eq.\ref{eq_rho_oo_new}, and integrating gives-
\begin{equation}
\dot \rho_{00}=\sum_{\sigma
\eta}\Gamma_{\sigma}^{\eta}[(1-f_{\eta}(\epsilon_{\sigma}))\rho_{\sigma\sigma}-f_{\eta}(\epsilon_{\sigma})\rho_{00}]
\end{equation}
Now in Ref.\cite{bingdong} the Fermi functions for the left and
right leads with respect to the electron spin
$f_{L}(\epsilon_{\uparrow})=f_{R}(\epsilon_{\uparrow})=1$ and
$f_{L}(\epsilon_{\downarrow})=f_{R}(\epsilon_{\downarrow})=0$.
Thus,
\begin{equation}
\dot\rho_{00}=-(\Gamma_{\uparrow}^{L}+\Gamma_{\uparrow}^{R})\rho_{00}+(\Gamma_{\downarrow}^{L}+\Gamma_{\downarrow}^{R})\rho_{\downarrow
\downarrow}
\end{equation}
Proceeding in exactly the same way, and using the Ref.\cite{thesis_pedersen} as a guide one can derive the other rate equations as written below. To model incoherence we turn to Ref.\cite{Kieblich} and use that as a model.

{\em Results}: We introduce density matrices $\rho_{ab}(t)$
meaning quantum dot is on the electronic state $|a>$
($a=b=0,\uparrow,\downarrow$) or on a  quantum superposition state
$(a\neq b)$ at time $t$. We introduce counting
fields\cite{bagrets}, $\chi_{\eta,\sigma}, \eta=L/R \mbox{ and }
\sigma=\uparrow/\downarrow$ to describe transitions from the dot
to leads. 

{\bf Coherent regime:} We first deal with the coherent regime.
$\dot{\rho}(t)=(\dot {\rho}_{00},\dot {\rho}_{\uparrow \uparrow},
\dot {\rho}_{\downarrow \downarrow},
{\Re(\dot\rho_{\uparrow\downarrow})},
{\Im(\dot\rho_{\uparrow\downarrow})})={\cal M} {\rho}(t)$, with
\begin{widetext}
\begin{equation}
\label{eq_M} {\cal M }=\left (
\begin{array}{ccccc}
-(\Gamma_{L\uparrow}+\Gamma_{R\uparrow})&0&(\Gamma_{L\downarrow}e^{i
\chi_{L\downarrow} }+\Gamma_{R\downarrow}e^{i \chi_{R
\downarrow}})&0&0\\
(\Gamma_{L\uparrow}e^{-i \chi_{L \uparrow}
}+\Gamma_{R\uparrow}e^{-i \chi_{R\uparrow}})&0&0&0&-2R_{rf}\\
0&0&-(\Gamma_{L \downarrow}+\Gamma_{R \downarrow})&0&2R_{rf}\\
0&0&0&-(\Gamma_{L\downarrow}+\Gamma_{R\downarrow})&-\delta_{ESR}\\
0&R_{rf}&-R_{rf}&\delta_{ESR}&-(\Gamma_{L\downarrow}+\Gamma_{R\downarrow})
\end{array}
\right),
\end{equation}
\end{widetext}
and $\delta_{ESR}=\Delta-\omega$. The normalization relation
$\rho_{00}+\sum_{\sigma \sigma}\rho_{\sigma\sigma}=1$ holds for
the conservation and $\Gamma_{\eta \sigma}=2\pi
\sum_{k}|V_{\eta}|^{2}\delta(w-\epsilon_{\eta k\sigma})$. We
assume the spin relaxation time of an excited spin state into the
thermal equilibrium to be very large.

We calculate the eigenvalues of Eq.\ref{eq_M}. The minimal of
these eigenvalues defines the full counting statistics (as,
$\chi_{\eta \sigma }\rightarrow 0, \eta=L,R;
\sigma=\uparrow,\downarrow$). After finding this eigenvalue
$Ev_{0}$, and then by using the approach pioneered in
Ref.\cite{bagrets}, We calculate the first, second and higher
cumulants. Note that the approach of Ref.\cite{bagrets} has been
generalized in Refs.\cite{Kieblich,sprekeler} to include both
coherent and incoherent transport regimes.
\begin{figure}[h]
\vskip 0.28in
\centerline{\includegraphics[width=8cm,height=6cm]{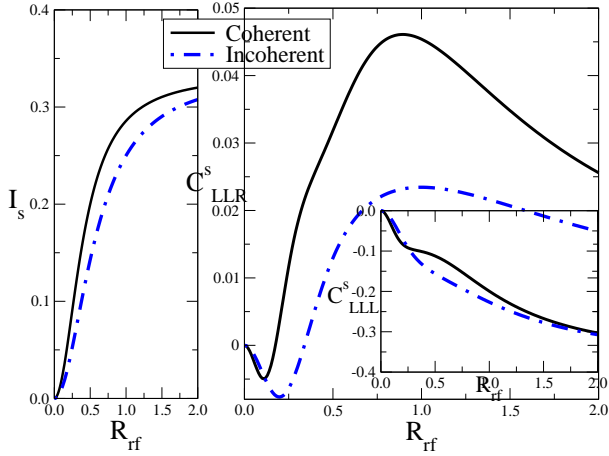}}
\caption{A comparison of coherent and incoherent transport
regimes. The odd moments- spin currents(left) and third moment
cross and auto-correlations. The parameters are $\Gamma=1,
\delta_{ESR}=0$.}
\end{figure}

\begin{figure}[h]
\vskip 0.28in
\centerline{\includegraphics[width=8cm,height=5cm]{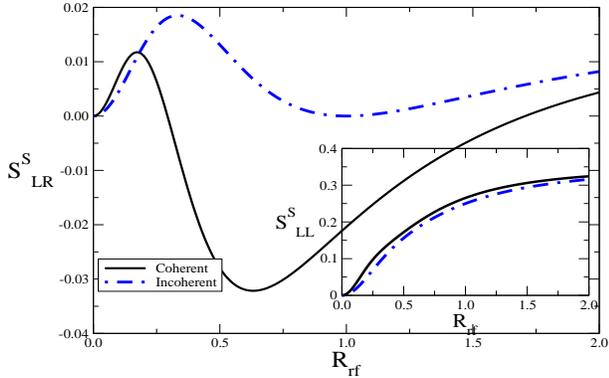}}
\caption{The second moment: The spin shot-noise cross-correlations. Parameters as mentioned before.}
\end{figure}

The first cumulant is defined as the current, we calculate the
individual spin polarized currents as follows: $I_{\eta
\sigma}=\frac{\partial Ev_{0}}{\partial \chi_{\eta
\sigma}}|_{\chi_{\eta \sigma}\rightarrow 0}$. The spin current is
thus $I^{s}_{\eta}=I_{\eta \uparrow}-I_{\eta \downarrow}$, while
the charge current is $I^{c}_{\eta}=I_{\eta \uparrow}+I_{\eta
\downarrow}$. The second cumulant defines the shot-noise. The shot
noise auto and cross-correlations can be calculated as follows.
The spin shot noise auto and cross-correlation is what we
concentrate on.
$S^{s}_{LL}=S^{\uparrow\uparrow}_{LL}+S^{\downarrow\downarrow}_{LL}-S^{\uparrow\downarrow}_{LL}-S^{\downarrow\uparrow}_{LL}$
and
$S^{s}_{LR}=S^{\uparrow\uparrow}_{LR}+S^{\downarrow\downarrow}_{LR}-S^{\uparrow\downarrow}_{LR}-S^{\downarrow\uparrow}_{LR}$
 wherein, $S^{\sigma \sigma'}_{\eta
\eta'}=\frac{\partial^{2}Ev_{0}}{\partial\chi_{\eta\sigma}\partial\chi_{\eta'\sigma'}}|_{\chi_{\eta
\sigma},\chi_{\eta' \sigma'}\rightarrow 0}$. Similarly the third
moment spin correlations are calculated as follows:
$C^{s}_{\eta\eta'\eta''}=C^{\uparrow \uparrow
\uparrow}_{\eta\eta'\eta''}+C^{\uparrow \downarrow
\downarrow}_{\eta\eta'\eta''}+ C^{\downarrow \uparrow
\downarrow}_{\eta\eta'\eta''}+C^{\downarrow \downarrow
\uparrow}_{\eta\eta'\eta''}-( C^{\uparrow \uparrow
\downarrow}_{\eta\eta'\eta''}+C^{\uparrow \downarrow
\uparrow}_{\eta\eta'\eta''} +C^{\downarrow \uparrow
\uparrow}_{\eta\eta'\eta''}+C^{\downarrow \downarrow
\downarrow}_{\eta\eta'\eta''})$. wherein  $C^{\sigma \sigma'
\sigma''}_{\eta
\eta'\eta''}=\frac{\partial^{3}Ev_{0}}{\partial\chi_{\eta\sigma}\partial\chi_{\eta'\sigma'}\partial\chi_{\eta''\sigma''}}|_{\chi_{\eta
\sigma},\chi_{\eta' \sigma'},\chi_{\eta'' \sigma''}\rightarrow
0}$. The existence of a pure spin current is a signature of a
spin-singlet electronic source. Since, in a pure spin current
electrons of opposite spin move in exactly opposite directions. 

{\bf Incoherent regime:}
To go into the incoherent or sequential transport regime as
exemplified in Refs.\cite{Kieblich}, we use the complete coherent
matrix, Eq.\ref{eq_M}, The coefficient matrix for incoherent
transport can be obtained from Eq.\ref{eq_M}, via setting
$\Re(\dot \rho_{\uparrow \downarrow})=0$ and $\Im(\dot
\rho_{\uparrow \downarrow})=0$ and then solving the two
simultaneous equations for $\Re(\rho_{\uparrow \downarrow}) $ and
$\Im(\rho_{\uparrow \downarrow})$ as in
Refs.\cite{Kieblich,sprekeler}. This leads to a $3X3$ matrix:
$\dot{\rho}(t)=(\dot {\rho}_{00},\dot {\rho}_{\uparrow \uparrow},
\dot {\rho}_{\downarrow \downarrow})={\cal M} {\bm \rho}(t)$ with
\begin{equation}\small{\cal M }=\left (
\begin{array}{ccccc}
-(\Gamma_{L\uparrow}+\Gamma_{R\uparrow})&0&\Gamma_{L\downarrow}e^{i
\chi_{L\downarrow} }+\Gamma_{R\downarrow}e^{i \chi_{R
\downarrow}}\\
\Gamma_{L\uparrow}e^{-i \chi_{L \uparrow}
}+\Gamma_{R\uparrow}e^{-i \chi_{R\uparrow}}&-z&z\\
0&z&-z-(\Gamma_{L \downarrow}+\Gamma_{R \downarrow})
\end{array}
\right),
\end{equation}
and,
$z=\frac{R^{2}_{rf}(\Gamma_{L\downarrow}+\Gamma_{R\downarrow})}{\delta^{2}_{ESR}+(\Gamma_{L\downarrow}+\Gamma_{R\downarrow})^{2}}$.
The minimal eigenvalue of this equation is again what we require.
$Ev_{0}=\frac{1}{6a}[K-2b-\frac{4(3ca-b^{2})}{K}]$, here
$K=36cba-108da^{2}-8b^{3}+12\sqrt{3}\sqrt{4c^{3}a-c^{2}b^{2}-18cbad+27d^{2}a^{2}+4db^{3}a}$,
and the elements $a,b,c,d$ are as follows (with
$\Gamma_{L\downarrow}=\Gamma_{L\uparrow}=\Gamma_{R\downarrow}=\Gamma_{R\uparrow}=\Gamma/2$):
\begin{eqnarray}
a&=&4\delta^{2}_{ESR}-\Gamma^{2}-\Gamma, b=8\Gamma\delta^{2}_{ESR}-8R^{2}_{rf}\Gamma-2\Gamma^{3}\nonumber\\
c&=&12R^{2}_{rf}\Gamma^{2}-\Gamma^{4}-4\Gamma^{2}\delta^{2}_{ESR}, X=R^{2}_{rf}\Gamma^{3}.\nonumber\\
d&=&4X-X(e^{-i\chi_{R\uparrow}}e^{i\chi_{R\downarrow}}+e^{-i\chi_{L\uparrow}}e^{i\chi_{R\downarrow}}\nonumber\\
&+&e^{-i\chi_{L\uparrow}}e^{i\chi_{L\downarrow}}+e^{-i\chi_{R\uparrow}}e^{i\chi_{L\downarrow}}).
\end{eqnarray}

In the incoherent regime too the spin current, spin shot-noise auto and cross correlations are calculated and finally the third moment auto and cross-correlations. In Fig. 2, the odd moments are plotted- pure spin current $I^{s}$ and the third moment auto $C_{3}=C^{s}_{LLL}$, and cross-correlations $C^{s}_{LLR}$. In Fig. 3, the second moment, shot noise auto $C_{2}=S^{s}_{LL}$ and cross-correlations $S^{s}_{LR}$. In all of these figures  the results for the coherent and incoherent transport regimes are contrasted. In both regimes the charge current is absolutely zero. Thus there is a pure spin current. The physics behind the pure spin current can be outlined as follows. In the model (Fig. 1) coulomb interaction in the quantum dot is strong enough to prohibit the double occupation, no more electrons can enter the quantum dot before the spin-down electron exits. As a result, the number of electrons exiting from the
quantum dot is equal to that of electrons entering the quantum dot; namely,
the charge currents exactly cancel out each other implying zero charge current.

{\em Conclusions:} The pure spin current obtained in our set-up is the spin-singlet electronic source which enables us to say that the positive noise cross-correlations obtained are a signature of the entangled state. Further we haven't put any noise dividers to and noise is generated by the currents in both the left and right leads obviating any source of doubt about any classical mechanism being responsible for positive cross-correlations. The main result of our work is depicted in Fig. 3, this is perhaps the first work where it is shown explicitly that the shot noise cross-correlations turn completely positive in the incoherent transport regime. What are the reasons for the completely positive shot noise cross-correlations? One can see from the formula for the spin shot noise cross-correlations it is a difference between same spin and opposite spin correlations. In the incoherent regime one notices that the magnitude of the same spin correlations, which are negative, is always less than that of the opposite spin case. In Fig. 2, the third moment auto and cross-correlations are also plotted. The impact of incoherence on the odd moment is distinctly muted as compared to that on the second moment. The third moment auto-correlations are completely negative as expected since the possibility of detecting three electrons is prohibited via Paulli exclusion. We have compared and contrasted the absolutely incoherent and absolutely coherent regimes. An effective parameter which shows the transition from completely coherent to completely incoherent can be introduced in the coherent density matrix, Eq. 16, to model this. Phenomenologically introducing a spin relaxation time into the coherent density matrix does indeed show the transition between completely coherent and incoherent regimes attesting our results.
\begin{table}[h]
\squeezetable
\begin{center}
\small
\caption{{Comparing first three moments in coherent and incoherent
regimes }}\vspace{0mm}
\begin{tabular}{|c|c|c|}
\hline
 Moment & {\em Coherent }& {\em Incoherent} \\ \hline
 $1^{st}$ & Pure spin current & Pure spin current \\ \hline
$2^{nd}$ & Shot-noise cross-correlations& Shot-noise\\
&positive for certain range &cross-correlations\\
 &of parameters & always positive.\\
 \hline
 $3^{rd}$ & Third moment finite& Third moment finite\\
 & & {\small No qualitative change} \\
 \hline
\end{tabular}
\end{center}
\end{table}
 In this letter, the first time the dramatic nature of
shot noise cross-correlations as a function of incoherence is shown. Future
endeavors on effects of incoherence on different geometries
especially including superconductors are contemplated.

\end{document}